\documentclass[%
reprint,
superscriptaddress,
amsmath,amssymb,
aps,
prl,showpacs
]{revtex4-1}

\usepackage{graphicx}% Include figure files
\usepackage{dcolumn}% Align table columns on decimal point
\usepackage{bm}% bold math
\usepackage{braket}
\usepackage{color}
\begin{document}

%\preprint{APS/123-QED}

\title{In-plane electric polarization of bilayer graphene nanoribbons by interlayer bias voltage
}% Force line breaks with \\

\author{Ryo Okugawa}
\affiliation{%
 Department of Physics, Tokyo Institute of Technology, 2-12-1 Ookayama, Meguro-ku, Tokyo 152-8551, Japan
}%
\author{Junya Tanaka}
\affiliation{%
 Department of Physics, Tokyo Institute of Technology, 2-12-1 Ookayama, Meguro-ku, Tokyo 152-8551, Japan
}%

\author{Takashi Koretsune}
\affiliation{%
RIKEN Center for Emergent Matter Science,
Hirosawa 2-1, Wako, Saitama 351-0198, Japan}%
\author{Susumu Saito}%
\affiliation{%
 Department of Physics, Tokyo Institute of Technology, 2-12-1 Ookayama, Meguro-ku, Tokyo 152-8551, Japan
}%
\affiliation{%
 TIES, Tokyo Institute of Technology, 2-12-1 Ookayama, Meguro-ku, Tokyo 152-8551, Japan
}%
\author{Shuichi Murakami}%
\affiliation{%
 Department of Physics, Tokyo Institute of Technology, 2-12-1 Ookayama, Meguro-ku, Tokyo 152-8551, Japan
}%
\affiliation{%
 TIES, Tokyo Institute of Technology, 2-12-1 Ookayama, Meguro-ku, Tokyo 152-8551, Japan
}%



\begin{abstract}
We theoretically show that an interlayer bias voltage in the AB-stacked bilayer 
graphene nanoribbons with armchair edges induces an electric
polarization along the ribbon. Both tight-binding 
and \textit{ab initio} calculations consistently
indicate that when the bias voltage is weak, the polarization shows
opposite signs depending on the ribbon width modulo three. This
nontrivial dependence is explained using a two-band effective model. A
strong limit of the bias voltage in the tight-binding model shows either one-third or zero
polarization, which agrees with topological argument.
%when the interlayer bias voltage is small enough compared to the hopping amplitude.
%From the analytically obtained polarization by the two-band model, we elucidate the %behavior of the polarization depending on the width in the weak bias voltage.
\end{abstract}

\maketitle

%\section{INTRODUCTIORY PARAGRAPH}

\begin{figure}%[b]
\includegraphics[width=8.5cm]{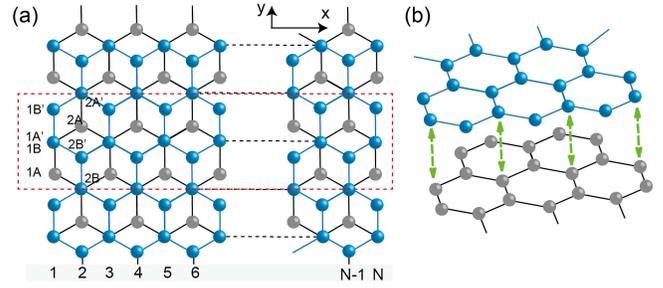}
\caption{\label{armgnr} (Color online) Structure of the AB-stacked bilayer GNR with the armchair edges.
(a),(b) The black (blue) lines represents the bondings in the lower (upper) layer. 
(a) $N$ is the width corresponding to a number of rows. The unit cell of the ribbons (red dashed-line box) contains $4N$ atoms.  
$n$A(A') and $n$B(B') represent the sublattice in the lower (upper) layer within $n$th
row.
(b) The green arrows denote the interlayer hoppings between the $n$B and $n$A' sites forming ``dimers''.
}
\end{figure}

 Monolayer graphene nanoribbons (GNRs) show various energy bands
depending on the edge orientation and the width of the nanoribbons\cite{fujita1996peculiar,PhysRevB.59.8271}.
When the GNRs have armchair edges, the energy bands become gapped or gapless, depending on the width.
Like monolayer GNRs, the energy bands
in the AB-stacked bilayer GNRs (Fig.~\ref{armgnr}) also depend sensitively 
on the edges.
Namely, the zigzag bilayer GNRs show localized edge states \cite{PhysRevLett.100.026802},
whereas the armchair bilayer GNRs vary from insulator ($N=3l$ or $3l+1$, $l$: integer) 
to metal ($N=3l+2$) by changing the width $N$  in a tight-binding (TB) model (Fig.~\ref{armgnr}) \cite{PhysRevB.78.045404}.  
In addition, external fields play scientifically and technologically important roles in atomic-layer materials, such as bilayer graphene. 
The external electric field opens up the fundamental gap in the 
AB-stacked bilayer graphene \cite{PhysRevLett.96.086805, PhysRevB.74.161403} which otherwise possesses massive and gapless parabolic bands in the low energy region \cite{PhysRevLett.96.086805}.

In this Letter, we theoretically show that an external interlayer bias voltage 
in the AB-stacked bilayer GNR with armchair edges induces 
a polarization along the ribbon direction. 
We use two methods: calculation on the TB model   
and \textit{ab initio} 
calculations. Both two methods consistently show that when the bias voltage is weak, the polarization shows a nontrivial dependence on the 
ribbon width, having opposite signs depending on the width modulo three.  
A strong limit of the bias voltage shows either one-third or zero polarization in the 
unit of the electron charge, which agrees with topological argument. 
We then discuss that the present theory applies to a wide variety of atomic-layer compounds. Thus nanostructure which breaks bulk symmetries allows novel responses which are absent in the bulk.

%The energy bands of the graphene have gapless linear dispersions near the Fermi energy, 
%called Dirac cones, because of its honeycomb structure.
%Furthermore, GNRs are known to 
%show remarkable electronic structure depending on the edges.
%For example, zigzag GNRs have flat-band edge states %\cite{fujita1996peculiar,PhysRevB.59.8271}.
%On the other hand, when a GNR has armchair edges,
%there is no edge state, and the one-dimensional bulk 
%energy bands become insulating or metallic, depending on the width of the nanoribbons \cite{fujita1996peculiar,PhysRevB.59.8271}. 
%In this paper, we discuss AB-stacked bilayer graphene (Fig.~\ref{armgnr}).

%The energy bands of the armchair GNRs are gapped when $N=3l$ or $3l+1$, where $l$ is an integer, while gapless when $N=3l+2$ \cite{PhysRevB.78.045404}.  

%\section{POLARIZATION INDUCED BY THE INTERLAYER BIAS}
%discussion on the symmetry
%In the present paper, we theoretically find that the AB-stacked bilayer GNRs show an electric polarization 
%along the ribbon direction when an interlayer bias voltage is applied.
We first discuss symmetry requirement for the transverse response
 of the polarization along the ribbon induced by the interlayer bias.
We take the $x$ and $z$ axis in the direction normal to the edges and the bilayer, 
respectively, and the $y$ axis along the ribbon (Fig.~\ref{armgnr}).
For armchair or chiral edges, when the interlayer bias voltage is zero,
inversion and $C_{2x}$ symmetries are preserved, which prohibit emergence of
polarization. Interlayer voltage breaks both symmetries, resulting in a
polarization along the ribbon. For zigzag edges, 
$xz$-plane mirror symmetry is preserved in addition to inversion symmetry,
and it prohibits emergence of polarization 
along the ribbon ($y$) direction. Because the interlayer voltage does 
not 
break this mirror symmetry, it does not induce polarization for zigzag ribbons. 

%polrization from TB model
First we numerically calculate the polarization for a spinless TB model.
\begin{equation}
H=\sum_{\langle i,j\rangle}t_{ij}c^{\dag}_ic_{j} + \frac{U}{2}\sum_{i}\xi _{i} c^{\dag}_ic_{i}.
\end{equation} 
The first term describes the hoppings with the amplitude $t_{ij}$, 
for which we only consider 
the nearest-neighbor intralayer hopping $t$ and the 
interlayer hopping  $t_\perp $ within a ``dimer''. 
Here, we set $t_\perp =0.13t$ $(t>0)$ according to Ref~\onlinecite{PhysRevB.75.155115}.
The second term represents the interlayer bias 
$U$ and $\xi _{i}$ takes $+1$ ($-1$) for the upper (lower) layers.

From the TB model, we calculate
the electronic contribution of the polarization 
$P$ in terms of the Berry connection within the modern theory of 
polarization \cite{PhysRevB.47.1651,resta1992theory,RevModPhys.66.899}.
It is calculated as a change of polarization $P(U)-P(0)$ by changing the interlayer 
bias voltage $U$. This calculation works only for insulators, and therefore 
we restrict ourselves to 
the insulating GNRs whose width is $N=3l$ or $3l+1$ 
($l$: integer).
Because $P(0)=0$ by inversion symmetry,
we obtain $P(U)$ numerically.

\begin{figure}[htb]
\includegraphics[width=8.5cm]{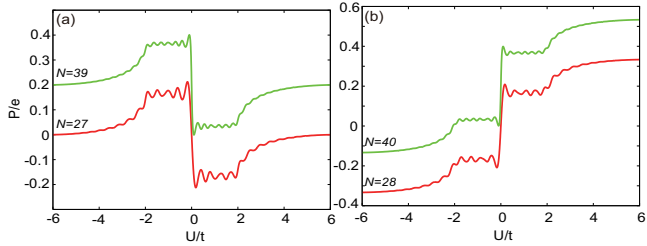}
\caption{\label{polar} (Color online) Numerical results of the polarization from the TB model in response to the interlayer bias.
The interlayer hopping parameter $t_{\perp}=0.13t$ is fixed.
(a) $N=3l$ ($N=27,39$) and 
(b) $N=3l+1$ ($N=28,40$).
The results for $N=39, 40$ (green) are vertically offset by 0.2.}
\end{figure}

Using this method, we find that in the bilayer GNRs with the armchair edges, 
the polarization arises for nonzero interlayer bias voltage $U$. 
Figures~\ref{polar}(a)(b) are our numerical results for various widths $N$.
In reality, feasible values of $U$ may be limited to about $|U|<0.15t$; nevertheless we show 
the results for much larger $U$ in the figure, to show consistency for the large $U$ limit. 
Interestingly, the behavior of the polarization is 
classified into two classes, $N=3l$ and $N=3l+1$. 
For $N=3l$, the polarization goes to 
zero for $U\rightarrow\pm\infty$, while for $N=3l+1$ the polarization 
goes to $\pm e/3$, where
$-e$ $(e>0)$ represents the electron charge. 
Furthermore, the slope around $U\sim 0$ has opposite signs between the two 
classes. The slope at $U\sim 0$ is steeper for wider ribbons. In the intermediate range 
of $U$ the polarization oscillates as $U$ is changed. This oscillation
accompanies a change of
band structure around the Fermi energy, 
formed by a number of minibands from a finite-size effect. 

\begin{figure}[htb]
\includegraphics[width=8cm]{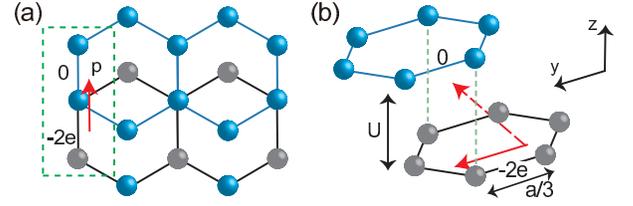}
\caption{\label{dipole} (Color online) Emergence of the electric dipole moment in the strong limit of the interlayer bias voltage $U$.
 (a) and (b) show top and side views of  a part of the nanoribbon.
$0$ and $-2e$ represent the charge at each layer for large $U$.
The dashed and solid red arrows represent the whole and the $y$ component of the dipole moment.
}
\end{figure}

%To elucidate this dependence on the width,
%we focus on the cases $U \rightarrow \pm \infty$ and $U\sim 0$.
%Firstly, we consider the polarization in  the limit $U \rightarrow \pm \infty$ from 
%Fig.~\ref{polar}:
%\begin{equation}
% P/e \sim \begin{cases}
%					0 & \text{$N=3l$} \\
%					\pm \frac{1}{3} & \text{$N=3l+1$}
%					\end{cases}.
%\end{equation}
The dependence of the asymptotic behavior at $U\rightarrow \pm\infty$ on the 
ribbon width can be physically understood as follows. 
Because there are two electrons per row in the unit cell,
two electrons lie on the lower layer under the strong interlayer bias voltage $U \sim +\infty$. 
As a result,  compared from $U\sim 0$, the two electrons are displaced by $a/6$ on average, and 
therefore 
the polarization is $P \equiv (a/6)\cdot 2Ne/a=Ne/3\ (\mathrm{mod}\ e)$ per unit length  (Fig.~\ref{dipole}), 
where
%$a$ is the distance between carbon atoms. 
$a$ is the lattice constant. 
Here the polarization is defined modulo $e$ \cite{PhysRevB.47.1651,resta1992theory,RevModPhys.66.899}.
Hence we obtain $P\equiv 0 \ (\mathrm{mod}\ e)$ and $P\equiv e/3 \ (\mathrm{mod}\ e)$
for $N=3l$ and $3l+1$, respectively. 
It totally agrees with our numerical calculations. 

\begin{figure}%[h]
\includegraphics[width=8.5cm]{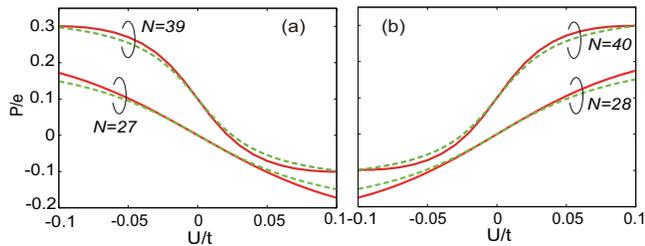}
\caption{\label{polaana} (Color online) Comparison between the results from 
the TB model (red solid line) and the 2B model (green dashed line).
(a) $N=3l$ ($N=27,39$) and 
(b) $N=3l+1$ ($N=28,40$).
The results for $N=39, 40$ are vertically offset by 0.1.}
\end{figure}

Next, we focus on the region of the weak interlayer bias voltage in  Fig.~\ref{polar}.
%Interestingly, our numerical calculationshows that the slope of polarization at $U=0$ is negative for $N=3l$ and positive for $N=3l+1$.
%In other words,
%\begin{equation}
%\frac{\partial  P}{\partial U}\Big| _{U=0}  \begin{cases}
%					 <0 & \text{$N=3l$} \\
%					>0 & \text{$N=3l+1$}
%					\end{cases}.
%\end{equation}
To understand the novel behavior of the slope, 
we construct a simple two-band (2B) effective Hamiltonian for the weakly biased GNRs, 
by retaining only the highest occupied band and the lowest unoccupied band.
To this end, we begin with the analytic 
forms of the eigenstates of the TB model at $k= 0$ and $U=0$
\cite{PhysRevB.78.045404}. 
The eigenvalue equation at $k= 0$, $U=0$ is written 
\begin{eqnarray}
\begin{pmatrix}
0 & t(2\cos \theta +1) \\
t(2\cos \theta +1) & \pm t_{\perp}
\end{pmatrix}
\begin{pmatrix}
A^{\pm} \\
B^{\pm}
\end{pmatrix}
=\varepsilon ^{\pm}
\begin{pmatrix}
A^{\pm} \\
B^{\pm}
\end{pmatrix},
%&&\varepsilon A^{\pm} = tB^{\pm}(1+2\cos \theta ), \\
%&&\varepsilon B^{\pm}=tA^{\pm}(1+2\cos \theta ) \pm t_{\perp}B^{\pm}.
\end{eqnarray}
where
$A^{\pm}$ corresponds to the sum and the difference between the amplitudes at 
the A and B' sublattices, respectively, and  $B^{\pm}$ is defined similarly for the B and A' sublattices (see the Supplemental Material \cite{Supplemental} for details).  
$\theta$ is a phase difference of an electronic wave
between the neighboring rows, forming a standing wave in the ribbon. 
Its eigenvalues are
\begin{eqnarray}
\varepsilon ^{\pm ,q}=\pm \frac{t_{\perp}}{2}+q\sqrt{\Bigl( \frac{t_{\perp}}{2}\Bigr) ^2+t^2(2\cos \theta +1)^2}, 
\end{eqnarray}
where $q=\pm 1$. From the boundary condition, we get $\theta=\theta^{N} _{r}=\frac{r}{N+1}\pi $, $r=1,2,\dots ,N$. 
Here, response of the 
polarization $ P$ to an external perturbation 
is given by the Berry curvature\cite{Berry84,PhysRevB.47.1651,RevModPhys.66.899}.
Therefore, the eigenstates close to $k=0$,  
where the band structure has a direct gap when $U\sim 0$, 
contributes considerably to the polarization. 
Hence, 
from the analytic forms of the eigenstates of the TB model at $k= 0$ and $U=0$
\cite{PhysRevB.78.045404}, we retain only the lowest unoccupied state $\ket{+}$ and the highest occupied state $\ket{-}$. Their energy eigenvalues are
given by $\pm g_0$, 
where $g_0=-t_{\perp}/2+d$, $d=\sqrt{(t_{\perp}/2)^2+t^2(2\cos \theta _{2l+1}^N +1)^2}$,
and $\theta _{2l+1}^N=  \pi (2l+1)/(N+1)$ for both  $N=3l$ and $N=3l+1$.
By using these two eigenstates, we construct a 2B 
model which describes the energy bands around the Fermi energy when $k\sim 0$ and $U \ll t,t_{\perp}$. 
The 2B Hamiltonian to the first order in $k$ and $U$ is derived as
\begin{equation}
H_{\rm eff}=h_1U\sigma _x+h_2k\sigma _y+g_0\sigma _z,
\end{equation} 
where $h_1=t_\perp /(4d), 
h_2=at^2(\cos \theta _{2l+1}^N-1)(2\cos \theta _{2l+1}^N+1)/(3d)$ and 
$\sigma _{x,y,z}$ are the Pauli matrices for the space spanned by the eigenstates $\ket{\pm}$ \cite{Supplemental}. 
%Detailed derivations are summarized in the Supplemental Material.
We note that width dependence appears through $\theta _{2l+1}^N$.
From the 2B Hamiltonian, we calculate the polarization $ P(U)$ for small $U$, 
as shown 
in Fig~\ref{polaana}. They well agree with the results of the TB model
in the $U\sim 0$ region, including the sign of the slope, for $N>20$. For $N<20$, 
the gap in the TB model is non-monotonous
as a function of the interlayer bias $U$; for small $U$ the gap decreases as a function of $U$. 
It is not reproduced in the 2B 
model where the gap always increases with $U$. This leads to differences 
between the two models.

%\begin{widetext}
%\begin{eqnarray}
% P(U)=-\frac{1}{2\pi}\arctan \Biggl( \frac{h_1h_2U\pi /a}{\varepsilon _{2l+1}^{-,+}\sqrt{({\varepsilon _{2l+1}^{-,+}}) ^2+h_1^ 2U^2+h_2 ^2\pi ^2/a^2}} \Biggr) , \\
%j(U)=-\frac{1}{2a}h_1h_2 \frac{\varepsilon _{2l+1}^{-,+}}{\bigl( h_1^2U^2+(\varepsilon _{2l+1}^{-,+})^2\bigr) \sqrt{h_1^2U^2+(\varepsilon _{2l+1}^{-,+})^2+h_2^2\pi ^2/a^2}}.
%&& P(U)=  \notag \\
%&&-\frac{1}{2\pi}\arctan \Biggl( \frac{h_1h_2U\pi /a}{g\sqrt{g^2+(h_1 U)^2+(h_2 \pi /a)^2}} \Biggr).
%&&j(U)=\\
%&&-\frac{1}{2a} \frac{h_1h_2g}{\bigl( (h_1U)^2+g^2\bigr) \sqrt{(h_1 U)^2+g^2+(h_2 \pi /a)^2}}. \label{effcur}
%\end{eqnarray}
%\end{widetext}
%This effective-model calculation well agrees with the numerical calculations 
%Therefore, we can explain the difference of the slopes for $N=3l$ and $N=3l+1$ from the effective model. 
The width dependence is understood from the analytic formula for the slope of $ P(U)$: \begin{eqnarray}
\frac{\partial  P}{\partial U}=\frac{ie}{2\pi} \int _{-\frac{\pi}{a}}^{\frac{\pi}{a}}dk \frac{\bra{u_{k-}}\frac{\partial H}{\partial k}\ket{u_{k+}}\bra{u_{k+}}\frac{\partial H}{\partial U}\ket{u_{k-}}}{(E_{k-}-E_{k+})^2}
+c.c.\ \  \label{current}
\end{eqnarray}
Here 
$E_{kj}$ and $\ket{u_{kj} }$ are the eigenvalue and the eigenstate of the 2B 
Hamiltonian for the $j$ band.
Therefore, the slope at $U=0$ is  \begin{eqnarray}
\frac{\partial  P}{\partial U}\Big| _{U=0}=-\frac{e}{2a}h_1h_2 \frac{1}{g_0 \sqrt{g_0^2+(h_2\pi /a)^2}}, \label{sign} 
\end{eqnarray}
and its sign is given by $-{\rm sgn}(h_2)$.
From $2\pi /3 -\theta _{2l+1}^N= \mp \pi /(3N+3)$ for $N=3l$ and $3l+1$, the sign of the 
slope is negative for $N=3l$ and positive for $N=3l+1$, 
in agreement with the results for various widths of GNRs.
%As a result, we find $\partial  P/\partial U |_{U=0}<0$ when $N=3l$ while $\partial  P/\partial U |_{U=0}>0$ when $N=3l+1$,
Asymptotic form for a wider ribbon is evaluated as
\begin{equation}
\frac{\partial  P}{\partial U}\Big| _{U=0}\sim\pm \frac{3t_{\perp}e}{8\pi^3 t^2}(N+1)^2.
\label{eq:ndep}
\end{equation}
where the signs $\pm$ is $-$ for $N=3l$ and $+$ for $N=3l+1$. 

These behaviors are confirmed by {\it ab initio} calculations based on density functional theory (DFT).
We perform the electronic structure calculation of hydrogen terminated AB-stacked bilayer GNRs within the local-density approximation (LDA)\cite{ceperley1980,perdew1981} based on DFT using {\sc Quantum Espresso} package \cite{Giannozzi}.
We use ultrasoft pseudopotentials\cite{vanderbilt1990} and plane-wave basis sets to describe the charge densities and wave functions with cutoff energies of 30Ry and 300Ry, respectively.
The supercell approach is used and the distances of neighboring bilayer GNRs along the $x$-axis and the $z$-axis are at least 10 and 30 \AA, respectively.
The geometries are fully optimized.
To discuss the effect of the external electric field, we apply a periodic zigzag potential along the $z$-axis in the supercell.
Under the external field, $E$, we obtain the band structure with $48 \times 1 \times 1$ k-points and calculate the electric polarization in terms of the Berry connection.

\begin{figure}%[h]
\includegraphics[width=8.7cm]{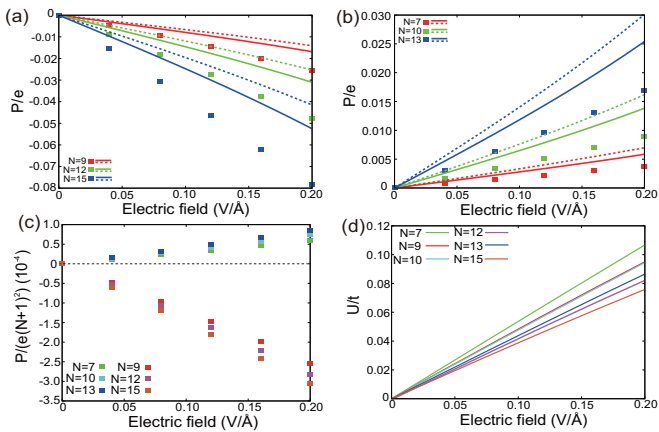}
\caption{\label{dft_polarization} (Color) 
Comparison between the results of DFT calculations and those with the TB model. 
(a), (b) Polarization obtained by DFT calculations (squares) compared with results with the TB model (lines).
The results of the TB with (without) rescaling 
by the gap size 
are shown as a solid line (dotted line) (see text).
(a) ($N=9,12,15$) are for 
 $N=3l$, and 
(b) ($N=7,10,13$) are for $N=3l+1$.
Note that it is for a spinless system, and the results here should be multiplied by two 
to compare with experiments. 
(c) Induced polarization $P$ by DFT calculation divided by $(N+1)^2$.
(d) Relationship between 
out-of-plane electric field $|{\bf E}|$ versus on-site potential energy difference $U$.
 $U$ is scaled by the hopping amplitude $t=2.6$eV \cite{PhysRevB.75.155115}.
}\end{figure}

The dot symbols in 
Fig.~\ref{dft_polarization} (a),(b) represent the DFT results on polarization for hydrogen terminated AB-stacked bilayer GNRs under the electric field $E$.
Here, we put $P(E=0)=0$ by symmetry.
Apparently, the signs of the slopes of $P$ obtained by DFT calculations completely agree with those from the TB and the 2B model.
Furthermore, the $N$ dependence of the polarization, $i.e.$, $\left. \partial  P(U)/\partial U\right|_{U=0} \propto (N+1)^2$ in Eq.~(\ref{eq:ndep}), is well reproduced in the DFT results
(see Fig.~\ref{dft_polarization} (c)).
These indicate that the simple TB model, and consequently its 2B effective model, well capture the key features of the polarization in this system.
To compare the polarization values with the TB model quantitatively, we relate the  electric field, $E$ to the on-site energies  $U$ in the TB model. To this end, 
we construct maximally localized Wannier
functions for the valence bands using carbon $\sigma$ and $\pi$ orbitals\cite{a,b,c}, and the result is shown in Fig.~\ref{dft_polarization} (d).
The obtained on-site energies for $\pi$ orbitals in each layer are almost independent of the position of the orbitals except at the edges.
Thus, in Fig.~\ref{dft_polarization} (d), we use average values excluding the edges.

As shown in Fig.~\ref{dft_polarization} (d),
we find that for weak electric field $|E|<0.2\mbox{V/\AA}$, $E$ and $U$ are almost linear,
 whereas their proportionality constant depends on the ribbon width. Using this correspondence
the results on the TB model over various $U$ is 
translated into the dependence on the electric field $E$, as shown as the
dotted lines in Fig.~\ref{dft_polarization} (a) and (b). We notice that the 
result from DFT and that from the TB models have similar tendencies, whereas
they are different by a factor of two smaller or larger, depending on the series 
$N=3l$ and $N=3l+1$. This difference between DFT and the TB model can be partly  attributed to the difference of the gap size. 
In the result of the 2B model in Eq.~(\ref{sign}), 
the polarization is inversely proportional to the gap size, because 
$h_2\pi /a$ is much larger than $g_0$ for the given parameters. Actually, the gap 
size obtained from DFT is smaller (larger) than that of the TB model for
$N=3l$ ($N=3l+1$) even at $U=0$. Therefore, in order to incorporate this 
difference of the gap size, we rescale the results of the 
polarization of the TB model 
by the ratio between the gaps from the DFT and that of the
TB model. This rescaling enhances (suppresses) the polarization for $N=3l$ ($N=3l+1$). 
After the rescaling, the results (solid lines in Fig.~\ref{dft_polarization} (a) and (b))
exhibits better
agreement with the DFT results. Thus
despite the simplicity of the TB model, it describes the various aspects of
the behavior of the 
polarization well including the width dependence. 
%This relationship can also be established from 
%the charge disproportionation between the bilayers, and the result remains 
%almost unchanged.
%We note that the ionic polarization 
%is less than $10^{-4}e$, much less than 
%the electronic contribution. 

%\begin{figure}%[h]
%\includegraphics[width=8.5cm]{pola3l_re2.eps}
%\includegraphics[width=8.5cm]{pola3lp1_re.eps}
%\includegraphics[width=8.5cm]{pola-141123.eps}
%\caption{\label{dft_polarization} \textbf{Polarization obtained by DFT calculations compared with results with the tight-bindng model.}
%(a) ($N=9,12,15$) are for 
% $N=3l$, and 
%(b) ($N=7,10,13$) are for $N=3l+1$.
%}
%\end{figure}

%\begin{figure}%[h]
%\includegraphics[width=8.5cm]{mod0p_2.eps}
%\includegraphics[width=8.5cm]{mod1p_2.eps}
%\caption{\label{dft_polarization} \textbf{Polarization obtained by DFT calculations.}
%(a) ($N=9,12,15$) are for 
% $N=3l$, and 
%(b) ($N=7,10,13$) are for $N=3l+1$.
%}
%\end{figure}

To experimentally measure this proposed effect, one needs a bilayer nanoribbon 
with well-defined edges and width.
For single-layer graphene nanoribbons, well-defined edge orientations have been demonstrated 
\cite{Magda,Han,Cai,Kosynkin}, and it might be realized also for bilayer graphene. 
For the bilayer 
graphene, interlayer electric field up to 0.3V/\AA\ has been achieved\cite{YuanboZhang}, and therefore 
the proposed effect with polarization up to $\sim -$0.12$e$ per spin for $N=15$ is expected to be realizable experimentally.
We also have calculated the effect of periodic modulations of the width
to check the edge disorder effect via supercell approach and 
% some types of weak disorders and
confirmed that the polarization survives the weak modulations considered \cite{Supplemental}.
Nevertheless, since the effect is sensitive to the ribbon width, the proposed effect will disappear 
in the presence of strong disorder.
We note here that the in-plane polarization by an interlayer bias can be 
expected for a wide variety of atomic-layer compounds, as long as the symmetry criterion 
for its emergence is satisfied. As an example, a bilayer armchair ribbon of transition metal dichalcogenides in the 2H stacking satisfies this criteria. Moreover, our calculation show induced polarization in AA'-stacked bilayer boron nitride nanoribbons \cite{Supplemental}. 
%Hence, the polarization is promising not only in the bilayer GNRs but also in various layered nanoribbons.
Such a wide choice of candidate materials 
provides us with many chances for experimental verifications of our theory.

To conclude, we theoretically show that the AB-stacked graphene nanoribbon 
with armchair edges has a polarization along the ribbon direction, when interlayer bias voltage
is applied. This is shown both by the simple tight-binding model and the {\it ab initio}
calculations. In particular, the linear response to the interlayer voltage shows different
signs for the cases $N=3l$ and $N=3l+1$, and it is fully understood by means of a simple
two-band model. 
%We evaluate the polarization for realistic strength of the interlayer electric field. 

\begin{acknowledgments}
This work is partially supported by Grant-in-Aid from MEXT, Japan (No. 26287062, No. 25107005 and No. 25104711), JSPS Research Fellowships for Young Scientists, and MEXT Elements Strategy Initiative to Form Core Research Center (TIES). 
\end{acknowledgments}

%\bibliographystyle{apsrev4-1}
%\bibliographystyle{naturemag}
%\bibliography{gp}% Produces the bibliography via BibTeX.

\renewcommand{\thefigure}{S\arabic{figure}}
\renewcommand{\theequation}{S\arabic{equation}}
\setcounter{equation}{0}
\setcounter{figure}{0}

\begin{widetext}

\section{Formula of the Polarization in terms of the Bloch wavefunctions}\label{pol1}
In our calculation of the polarization $ P$ in terms of the Bloch wavefunctions, 
we used the formula 
\cite{resta1992theory, PhysRevB.47.1651,RevModPhys.66.899}
\begin{equation}
 P=
-\frac{ie}{2\pi} \int _{-\frac{\pi}{a}}^{\frac{\pi}{a}}dk \sum_{n}^{\rm occ.}
\bra{u_{kn}}\frac{\partial }{\partial k}\ket{u_{kn}},
\label{pola1}
\end{equation}
where $\ket{u_{kn}}$ is the Bloch wavefunction satisfying the cell-periodic gauge 
condition
\begin{equation}
u_{k,n}(\bm{r})=e^{iGy}u_{k+G,n}(\bm{r}),
\end{equation}
and the summation is taken over the occupied states below the Fermi 
energy.
Here $k$ is the Bloch wavenumber and $G\equiv 2\pi/a$ is a reciprocal lattice vector.
In numerical calculation, the differentiation in terms of $k$ in Eq.~(\ref{pola1}) should
be replaced by a difference in $k$. Such a formula with this replacement is discussed in detail in   
Ref.~\onlinecite{PhysRevB.47.1651}, and we followed this formalism for the calculation of the 
polarization.

\section{Polarization calculated from
 the tight-binding model for various widths}\label{sec:polaN}
We show numerical results of the polarization 
induced by the interlayer bias for graphene nanoribbons (GNRs) with various widths $N$, calculated from the tight-binding model. Some results are shown in Fig. 2 in the main text, and 
 we show more examples for various $N$ in Fig.~\ref{polaN}.
For $N=3l$ the results are shown in Fig.~\ref{polaN}{\bf a}, {\bf c} ($N=9,15,21$),
and in Fig. 2{\bf a}, {\bf b} ($N=27,39$).
For $N=3l+1$ the results are shown in Fig.~\ref{polaN}{\bf b}, {\bf d} ($N=10,16,22$),
and in Fig. 2{\bf c}, {\bf d} ($N=28,40$).
In the wide range of $U$ (Fi.g~\ref{polaN}), 
the polarization $P$ oscillates as a function of $U$, which
is attributed to crossings of minigaps. There are more oscillations for larger $N$, 
which reflects the fact that there are a larger number of minibands for wider ribbons. 
On the other hand, the oscillation amplitude becomes gradually smaller for wider ribbons
(see Fig. 2) because the contribution from each miniband becomes relatively smaller. 
On the other hand, in the regime $U\sim 0$, we showed from the two-band model
that 
the polarization is linear in $U$ with its slope scales with $(N+1)^2$. This is roughly 
reproduced for the results shown in Fig.~\ref{polaN}{\bf c} and {\bf d}.
\begin{figure}[h]
\includegraphics[width=10cm]{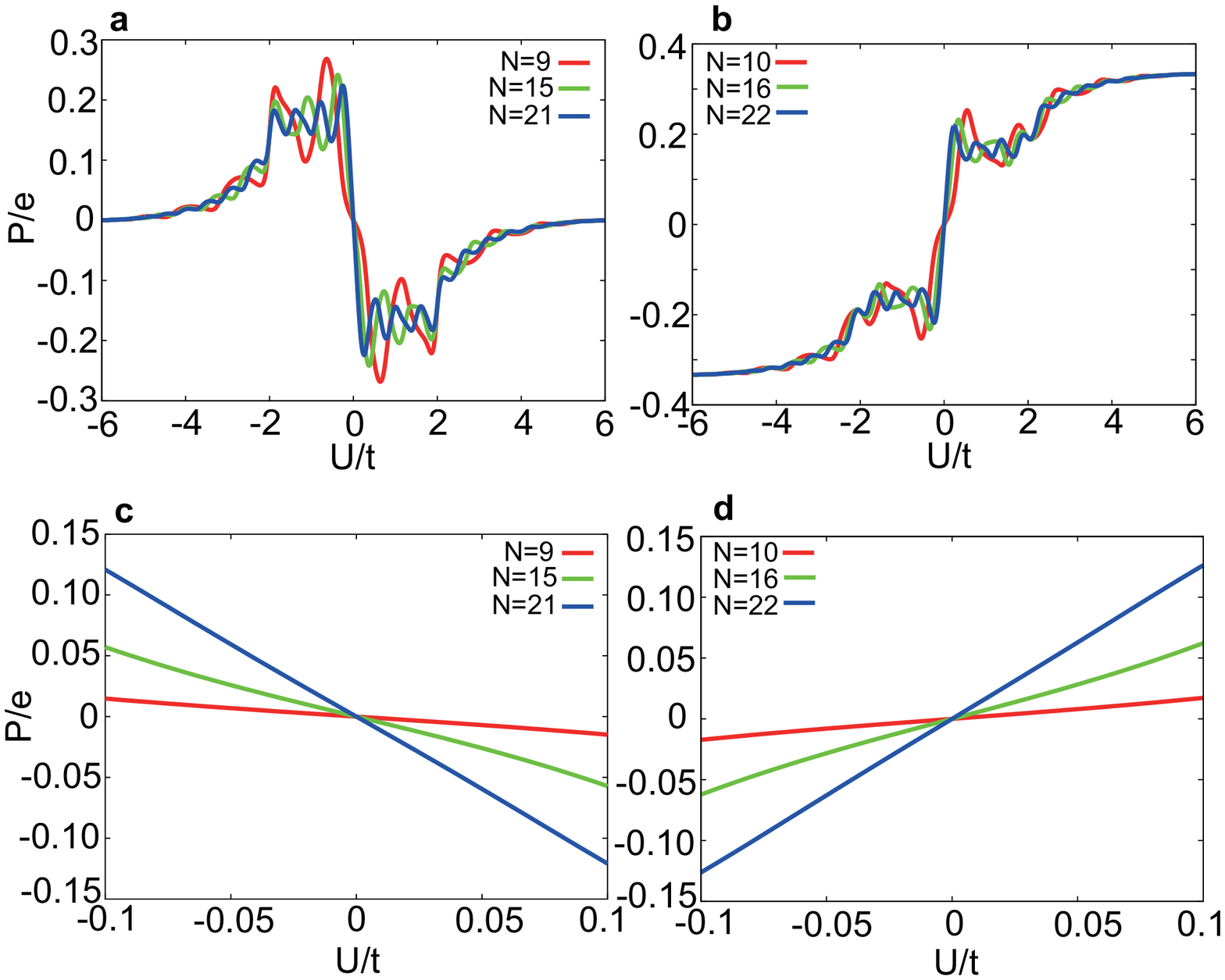}
\caption{\label{polaN} \textbf{Polarization induced by the interlayer bias $U$ for GNRs with various widths $N$, calculated from the tight-binding model.}
It is shown as a function of the interlayer bias $U$. 
\textbf{a} and \textbf{c} are for the class $N=3l$ and
\textbf{b} and \textbf{d} are for the class $N=3l+1$. 
}
\end{figure}

It may look strange that for a large $N$ limit, i.e. the 2D graphene limit,
the polarization has a different asymptotics for $N=3l$ and $N=3l+1$. 
It is in fact reasonable; the polarization per area is proportional to $P$ divided by the width, 
and therefore goes to zero for $N\rightarrow\infty$.

\section{Eigenvalues and eigenvectors of the tight-binding model}\label{eigen}
Here we derive the two-band low-energy effective Hamiltonian from the eigenstates of the tight-binding model of the bilayer GNR with armchair edge, without the interlayer bias voltage:
\begin{equation}
H_t=\sum_{<i,j>}t_{ij}c^{\dag}_ic_{j}.
\end{equation} 
Firstly, to obtain the two-band Hamiltonian, we diagonalize the tight-binding model at $k=0$ according to Ref.~\onlinecite{PhysRevB.78.045404}.
We set the eigenvector of the Hamiltonian 
\begin{eqnarray}
\ket{\Phi (k=0)}=\sum_{m=1}^{N}a_m\ket{mA}+\sum_{m=1}^{N}b_m\ket{mB}+\sum_{m=1}^{N}a_m'\ket{mA'}+\sum_{m=1}^{N}b_m'\ket{mB'},
\end{eqnarray} 
where $A$ and $B$ represent sublattices in the lower layer, and $A'$ and $B'$ in the upper layer.
Here, $a_m$, $b_m$, $a_m'$, and $b_m'$ are expansion coefficients.
From the tight-binding model at $k=0$, we obtain
\begin{eqnarray}
&&\varepsilon a_m = t(b_{m-1}+b_{m}+b_{m+1}), \label{e1} \\
&&\varepsilon b_m = t(a_{m-1}+a_{m}+a_{m+1})+t_{\perp}a_m', \\
&&\varepsilon a_m' = t(b_{m-1}'+b_{m}'+b_{m+1}')+t_{\perp}b_m, \\
&&\varepsilon b_m' = t(a_{m-1}'+a_{m}'+a_{m+1}'), \label{e4}
\end{eqnarray}
where $\varepsilon$ represents the energy eigenvalue.
To solve the above equations (\ref{e1})-(\ref{e4}), we introduce new coefficients
\begin{equation}
\alpha _m^{\pm} = \frac{1}{\sqrt{2}}(a_m\pm b_m'),\
\beta _m^{\pm} = \frac{1}{\sqrt{2}}(b_m\pm a_m').
\end{equation}
Then, we obtain
\begin{eqnarray}
&&\varepsilon \alpha _m^{\pm}=t(\beta _{m-1}^{\pm} +\beta _{m}^{\pm}+\beta _{m+1}^{\pm}), \label{z1} \\
&&\varepsilon \beta _m^{\pm}=t(\alpha _{m-1}^{\pm} +\alpha _{m}^{\pm}+\alpha _{m+1}^{\pm}) \pm t_{\perp} \beta _{m}^{\pm}. \label{z2}
\end{eqnarray}
Since $\alpha_0^{\pm}=0=\beta^{\pm}_0$, the solutions have the 
form
 $\alpha _m^{\pm} \propto A^{\pm}\sin(m\theta)$ and $\beta _m^{\pm} \propto B^{\pm}\sin(m\theta)$, where $\theta$ is a constant $(0<\theta<\pi)$. We rewrite the equations (\ref{z1}) and (\ref{z2}) in the matrix form
\begin{eqnarray}
\begin{pmatrix}
0 & t(2\cos \theta +1) \\
t(2\cos \theta +1) & \pm t_{\perp}
\end{pmatrix}
\begin{pmatrix}
A^{\pm} \\
B^{\pm}
\end{pmatrix}
=\varepsilon ^{\pm}
\begin{pmatrix}
A^{\pm} \\
B^{\pm}
\end{pmatrix},
%&&\varepsilon A^{\pm} = tB^{\pm}(1+2\cos \theta ), \\
%&&\varepsilon B^{\pm}=tA^{\pm}(1+2\cos \theta ) \pm t_{\perp}B^{\pm}.
\end{eqnarray}
and its eigenvalues are obtained analytically.
\begin{eqnarray}
\varepsilon ^{\pm ,q}=\pm \frac{t_{\perp}}{2}+q\sqrt{\Bigl( \frac{t_{\perp}}{2}\Bigr) ^2+t^2(\cos \theta +1)^2}, 
\end{eqnarray}
where $q=\pm 1$.
From the boundary condition,
the coefficients must vanish when $m=0$ and $N+1$ and we get
\begin{equation}
\theta _{r}=\frac{r}{N+1}\pi , \hspace{3mm} \text{$r=1,2,\dots ,N$}.
\end{equation}
Thus, we get $4N$ eigenvalues and eigenvectors,
\begin{eqnarray}
%\varepsilon _{r}^{\pm ,q}=\pm \frac{t_{\perp}}{2}+q\sqrt{\Bigl( \frac{t_{\perp}}{2}\Bigr) ^2+t^2(\cos \theta _r+1)^2}, \\
\varepsilon _{r}^{\pm ,q}&&=\pm \frac{t_{\perp}}{2}+qd_r, \label{energy} \\
%\ket{r,\pm,q}
\begin{pmatrix}
A^{\pm,q}_r \\
B^{\pm,q}_r
\end{pmatrix}
&&=\frac{1}{\sqrt{2d_r(d_r\mp q\frac{t_{\perp}}{2}})}
\begin{pmatrix}
\mp \frac{t_{\perp}}{2}+qd_r \\
t(2\cos \theta _r +1)
\end{pmatrix}.
\end{eqnarray}
We put $d_r=\sqrt{(t_{\perp}/2)^2+t^2(2\cos \theta _r+1)^2}$ for notational simplicity.
These energy eigenvalues can become zero only when $\cos \theta _r = -1/2$.
When $N=3l+2$ it can be satisfied for $r=2l+2$, and the energy bands are gapless at $k=0$ because $\varepsilon ^{-,+}_{2l+2}=\varepsilon ^{+,-}_{2l+2}=0$.
On the other hand, when $N=3l$ or $3l+1$, 
$\cos \theta _r = -1/2$ cannot be satisfied 
the energy bands are gapped. 
Therefore, the polarization $P_y(U)$ can be defined in the nanoribbons with widths $N=3l$ or $3l+1$.
In these cases, the energy eigenvalues closest to zero are $\varepsilon ^{-,+}_{2l+1}$ and $\varepsilon ^{+,-}_{2l+1}$, and the corresponding eigenstates $\ket{+}\equiv
\ket{r=2l+1, -,+}$ and $\ket{-}\equiv \ket{r=2l+1,+,-}$ considerably contribute to the polarization.
For brevity, we write $\pm g_0=\varepsilon ^{\mp, \pm}_{2l+1}$, and 
\begin{eqnarray}
g_0=&&-\frac{t_{\perp}}{2}+d_{2l+1}, \\
\ket{\pm }=&&\frac{1}{\sqrt{2\sum _{m=1}^N\sin ^2 (m\theta _{2l+1})}}\times \nonumber
\\
&& \sum _{m=1}^N \sin (m\theta_{2l+1})\Bigl(A_{2l+1}^{\mp ,\pm}(
\ket{mA}\mp\ket{mB'} ) +
B_{2l+1}^{\mp ,\pm}(\ket{mB}\mp\ket{mA'}) \Bigr) .
\end{eqnarray}
Hereafter, we omit subscripts $2l+1$ except for that in $\theta _{2l+1}$; for example, 
we write $d=d_{2l+1}$.

\section{Construction of the two-band Hamilotnian}\label{effH}
To elucidate the polarization in the weak interlayer bias voltage, we construct a two-band low-energy effective Hamiltonian for the space spanned by $\ket{\pm}$.
Therefore, we add the interlayer bias voltage $U$ to the tight-binding model as a perturbation;
\begin{eqnarray}
H_U=\frac{U}{2}\sum_{i}\xi _{i} c^{\dag}_ic_{i},
\end{eqnarray}
We retain terms up to the linear order in $k$ and the nonzero matrix elements to this order are given by
\begin{eqnarray}
\bra{mA}H_t\ket{nB}=&&\bra{mA'}H_t\ket{nB'} \nonumber \\
=&&t(\delta _{m,n-1}+\delta _{m,n}+\delta _{m,n+1}) +ikt\frac{a}{3}(\delta_{m,n}-\frac{1}{2}\delta _{m-1,n}-\frac{1}{2}\delta _{m+1,n}), \\
\bra{mA'}H_t\ket{nB}=&&t_{\perp}\delta _{m,n}, \\
\bra{mA}H_U\ket{nA}=&&\bra{mB}H_U\ket{nB}=-\frac{U}{2}\delta _{m,n}, \\
\bra{mA'}H_U\ket{nA'}=&&\bra{mB'}H_U\ket{nB'}=\frac{U}{2}\delta _{m,n}.
\end{eqnarray}
Therefore, the nonzero matrix elements of $H_t$ and $H_U$ are
\begin{eqnarray}
\bra{\pm }H_t\ket{\pm }=&&\pm g_0, \\
\bra{+}H_t\ket{-}=&&ikt^2\frac{a}{3d}(1-\cos \theta _{2l+1})(2\cos \theta _{2l+1}+1), \\
\bra{+}H_U\ket{-}=&&\frac{Ut_\perp}{4d}.
\end{eqnarray}
Hence, we obtain the two-band Hamiltonian  $H_{\rm eff}=H_t+H_U$,
\begin{equation}
H_{\rm eff}(k,U)=h_1U\sigma _x +h_2k \sigma _y +g_0\sigma _z,
\end{equation}
where
\begin{equation}
h_1=\frac{t_{\perp}}{4d},\ 
h_2=-t^2\frac{a}{3d}(1-\cos \theta _{2l+1})(2\cos \theta _{2l+1}+1).
\end{equation}
The eigenvalues and eigenstates of this two-band Hamiltonian are
\begin{eqnarray}
E_{k,\pm}&&=\pm \sqrt{(h_1U)^2+(h_2k)^2+g_0^2} =\pm g, \\
\ket{u_{k,\pm}}&&=\frac{1}{\sqrt{2g(g\pm g_0)}}
\begin{pmatrix}
g_0\pm g \\
h_1U+ih_2k
\end{pmatrix}.
\end{eqnarray}

\section{Polarization from the two-band Hamiltonian}\label{pol}
We focus on  the region of the weak interlayer bias voltage to clarify the difference of the slope of the polarization at $U \sim 0$ for two classes $N=3l$ and $N=3l+1$.  
The polarization $ P$ is given by \cite{resta1992theory, PhysRevB.47.1651,RevModPhys.66.899}
\begin{equation}
 P=\int_{0}^{U} dU' j(U'), \label{pola}
\end{equation}
where
%\begin{widetext}
\begin{eqnarray}
j(U)=\frac{ie}{2\pi} \int _{-\frac{\pi}{a}}^{\frac{\pi}{a}}dk \sum_{n}^{\rm occ.}\sum_{m}^{\rm unocc.} 
\frac{\bra{u_{kn}}\frac{\partial H}{\partial k}\ket{u_{km}}\bra{u_{km}}\frac{\partial H}{\partial U}\ket{u_{kn}}}{(E_{kn}-E_{km})^2}
+c.c. \label{current}
%j(U)=-\textrm{Im} \Bigl( \frac{e}{\pi}\int _{-\frac{\pi}{a}}^{\frac{\pi}{a}}dk \sum_{n}^{occ.}\sum_{m}^{unocc.} \times \notag \\
%\frac{\bra{u_{kn}}\frac{\partial H}{\partial k}\ket{u_{km}}\bra{u_{km}}\frac{\partial H}{\partial U}\ket{u_{kn}}}{(E_{kn}-E_{km})^2} \Bigr) . \label{current}
\end{eqnarray}
%\end{widetext}
Here, $j(U)=\partial  P/\partial U$, and $n$ and $m$ are band indices for
occupied bands and unoccupied bands, respectively.
$E_{kj}$ and $\ket{u_{kj} }$ are the $j$th eigenvalue and the eigenstate of the Hamiltonian.
In the present case $n=-$ and $m=+$, and therefore
$ P$ and $j(U)$ are calculated by using the two-band Hamiltonian derived in section~\ref{effH}.
\begin{eqnarray}
 P(U)=-\frac{e}{2\pi}\arctan \Biggl( \frac{h_1h_2U\pi /a}{g_0\sqrt{g_0^2+(h_1 U)^2+(h_2 \pi /a)^2}} \Biggr) , \\
j(U)=-\frac{e}{2a}h_1h_2 \frac{g_0}{\bigl( (h_1U)^2+g_0^2\bigr) \sqrt{(h_1 U)^2+g_0^2+(h_2 \pi /a)^2}}. \label{effcur}
\end{eqnarray}
In particular, the slope of the polarization at $U=0$, i.e. $j(0)$, is given by
\begin{equation}
j(0)=-\frac{e}{2a}h_1h_2\frac{1}{g_0\sqrt{g_0^2+(h_2\pi /a)^2}}.
\label{j0}
\end{equation}
Because $t_{\perp}>0$, $h_1$ and $g_0$ are positive, and we have
\begin{equation}
{\rm sgn}(j(0))=-{\rm sgn}(h_2)={\rm sgn}\Bigl( \frac{2}{3}\pi -\theta _{2l+1}\Bigr).
\end{equation}
Here,
\begin{eqnarray}
\frac{2}{3}\pi -\theta _{2l+1}=\begin{cases}
					-\frac{\pi}{3(N+1)} & \text{$N=3l$} \\
					 \frac{\pi}{3(N+1)} & \text{$N=3l+1$}
					\end{cases}.
\label{23pitheta}
\end{eqnarray}
Therefore, the sign of  $j(0)=\partial  P/\partial U |_{U=0}$ is given by 
\begin{equation}
 \frac{\partial  P}{\partial U}\Big| _{U=0} \begin{cases}
					 <0 & \text{$N=3l$} \\
					>0 & \text{$N=3l+1$}
					\end{cases},
\end{equation}
which well agrees with the numerical results of the tight-binding model.
Asymptotic behavior of $j(0)$ (Eq.~(\ref{j0})) is evaluated for large $N$, where 
$g_0\ll h_2\pi/a$;
\begin{equation}
j(0)=-{\rm sgn}(h_2)\frac{h_1e}{2\pi g_0}\sim\pm \frac{t_{\perp}e}{8\pi d(d-\frac{t_{\perp}}{2})}.
\end{equation}
where $\pm$ is $-$ for $N=3l$ and $+$ for $N=3l+1$. By using equation (\ref{23pitheta}) 
it is approximated as
\begin{equation}
j(0)\sim\pm \frac{3t_{\perp}e}{8\pi^3 t^2}(N+1)^2.
\end{equation}

\section{Effect of periodic modulation of the ribbon width}
We numerically calculate the polarization in bilayer GNRs with weak periodic modulations
of  the ribbon widths by using the tight-binding model
when the interlayer bias voltage is weak.
We consider two cases of periodic modulations 
by changing the width of each layer in various ways.

Firstly, we discuss an effect of the difference of the widths of the upper and lower layers.
Figure \ref{NLNU} shows the polarization in the armchair bilayer GNRs when the upper and lower layers have the different widths. 
Then, we calculate the polarization by changing the width for the upper layer 
$N_{\rm U}$ while fixing that for the lower layer $N_{\rm L}$,
except that $N_{\rm L}=2$ or $N_{\rm U}=2$ (mod 3) since the energy bands are gapless at $U=0$.
When $N_{\rm L}=N_{\rm U}$, the system is the perfectly stacked bilayer armchair GNRs in Fig.~1.
The results in Fig.~\ref{NLNU} \textbf{b} and \textbf{c}
correspond to the polarization for $N_{\rm L}=12\equiv 0$ and $N_{\rm L}=13\equiv 1$ (mod 3), respectively.
Consequently, we can see that all the slopes of polarization in Fig.~\ref{NLNU} have the same sign regardless of the width of the upper layer $N_{\rm U}$. 
%Namely, in this case, the sign of the slope of the polarization depends on only the width of the narrow layer.
Furthermore, even if the bilayer is composed of two layers with the width 0 and 1 (mod 3),
we find that the sign of the slope is equal to that of the perfectly stacked bilayer GNRs with the narrower width although the magnitude becomes small.
Therefore, when the widths of the upper and lower layers are different,
the polarization behaves like the perfectly stacked armchair GNRs with the width $\min (N_{\rm U}, N_{\rm L})$. 
%The decrease of the polarization can be understood from the opposite slopes of the polarization of the different widths. 

\begin{figure}[h]
\includegraphics[width=12cm]{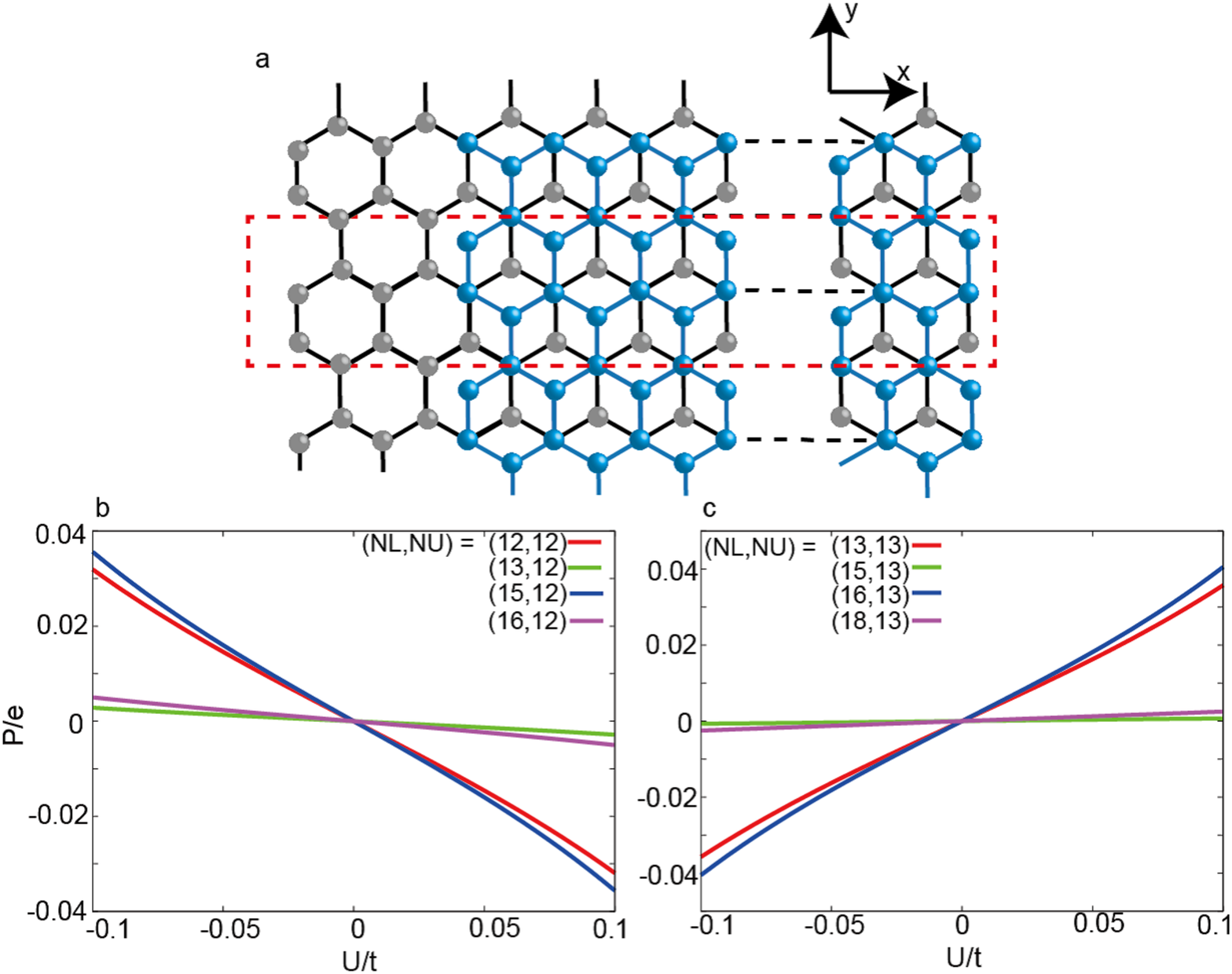}
\caption{\label{NLNU} \textbf{Polarization in the bilayer composed of two layers with different widths.}
\textbf{a} shows structure of the GNRs with the different upper and lower layer. 
The unit cell is described by the red dashed-line box.
\textbf{b} and \textbf{c} show the polarization for $N_{\rm U}=12$ and $N_{\rm U}=13$, respectively.
\textbf{b} is for $N_{\rm U}=3l$, and \textbf{c} is for $N_{\rm U}=3l+1$.  
}
\end{figure}

Secondly, we calculate the polarization of bilayer GNRs when the widths of 
the upper and lower layers alternates between two values  $N_1$ and $N_2$ $(N_1\geq N_2)$, as shown in Fig.~\ref{N_1N_2} \textbf{a} having no dangling bonds.
In this system, the primitive translation vector doubles.
In particular, when $N_1=N_2$, the system corresponds to the armchair bilayer GNRs. 
%do not take account into dangling bonds because in general they are unstable chemically.
In this case, when we calculate the polarization, we change only $N_1$ and fix $N_2$.
The results of the polarization for various widths $(N_1, N_2)$ are shown in Fig.~\ref{N_1N_2}.
Figure \ref{N_1N_2} \textbf{b} and \textbf{c} show the polarization for $N_2=12\equiv 0$ and $N_2=13\equiv 1$ (mod 3), respectively.
As a result,  as $N_1$ becomes larger, we find that the magnitudes of the polarization for $N_2\equiv 0 $ (mod 3) are enhanced, while those for  $N_2\equiv 0 $ (mod 3) are suppressed. 
Nevertheless, the sign of the slope of the polarization is unchanged from that of the perfect armchair GNRs with the width $N_1=N_2$.
Therefore, we can see that the polarization in the weak interlayer bias voltage
is dominated by the narrow part of GNRs, which is similar to the previous case. 

\begin{figure}[h]
\includegraphics[width=12cm]{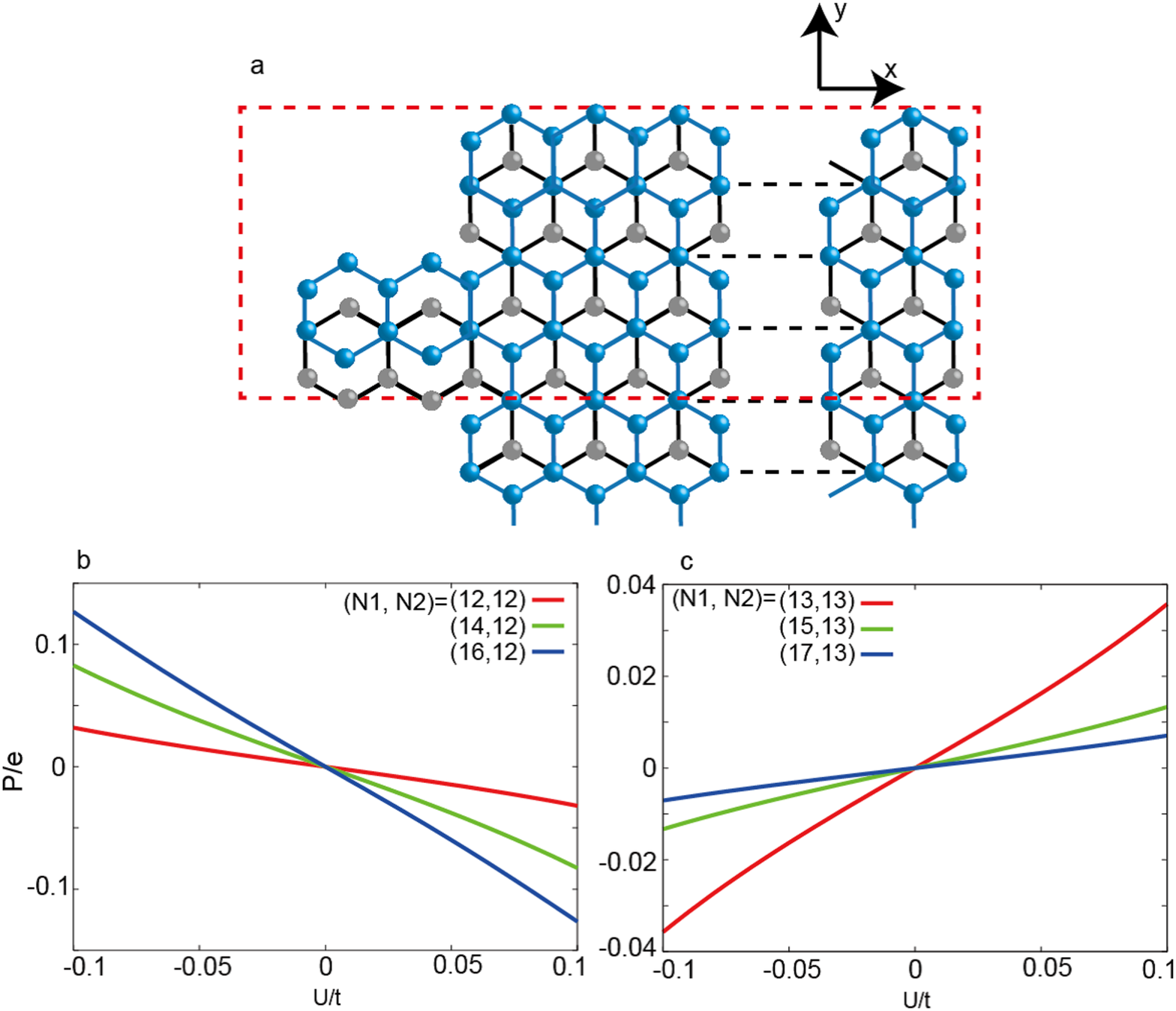}
\caption{\label{N_1N_2} \textbf{Polarization in the bilayer consisting of the stacked monolayer with two widths.}
\textbf{a} shows the structure of the bilayer nanoribbons with two widths.
The red dashed-line box describes the unit cell.
\textbf{b} and \textbf{c} are examples of the polarization for $N_{\rm U}=12$ and $N_{\rm U}=13$, respectively.
\textbf{b} is for $N_{\rm U}=3l$, and \textbf{c} is for $N_{\rm U}=3l+1$.  
}
\end{figure}

From the above results, we find that small variations of the width do not affect the sign of the slope of the polarization in the weak interlayer bias voltage.
In other words, even though the edges of the bilayer graphene nanoribbons are not completely perfect, 
the nontrivial dependence of the polarization on the width appear like the perfectly stacked bilayer GNRs with armchair edges. 
Thus, the in-plane polarization in response to the interlayer voltage survives even if the edges have weak periodic modulation of the ribbon width, as long as the energy bands are gapped.
Nevertheless, since the effect is sensitive to ribbon width, the proposed effect will disappear 
in the presence of strong disorder.

\section{Polarization in the bilayer BN nanoribbon}
We explained the emergence of the polarization in bilayer graphene nanoribbons 
by symmetry argument.
Therefore, other nanoribbons of atomic-layer compounds 
can have a finite polarization along the ribbon direction in response to the interlayer voltage, when the symmetry criterion for such a response 
is satisfied. 
To confirm this, we compute the polarization of hydrogen terminated bilayer BN nanoribbons from first-principles calculations.
Here, we consider so-called AA'-stacked bilayer BN nanoribbons with the armchair edges and the geometry is fully optimized (see Fig.~\ref{BNNR_polarization} (a)).
As in the case of the bilayer armchair GNRs, 
$xz$-mirror symmetries are broken in this structure.
When the interlayer bias voltage is zero, inversion symmetry is preserved and the polarization 
is zero. The interlayer voltage breaks the inversion symmetry, leading to nonzero polarization 
along the ribbon, as we see in the following. 
Figure \ref{BNNR_polarization} (b) shows calculated in-plane polarization as a function of the interlayer voltage.
As we can see, a finite polarization appears as expected.
Note that the size of the polarization is rather small compared to those in the GNRs.
One reason is that the band gaps of these nanoribbons are relatively large ($\sim 4.3$ eV both for $N=7$ and $N=9$ at $U=0$).

\begin{figure}[h]
\includegraphics[width=14cm]{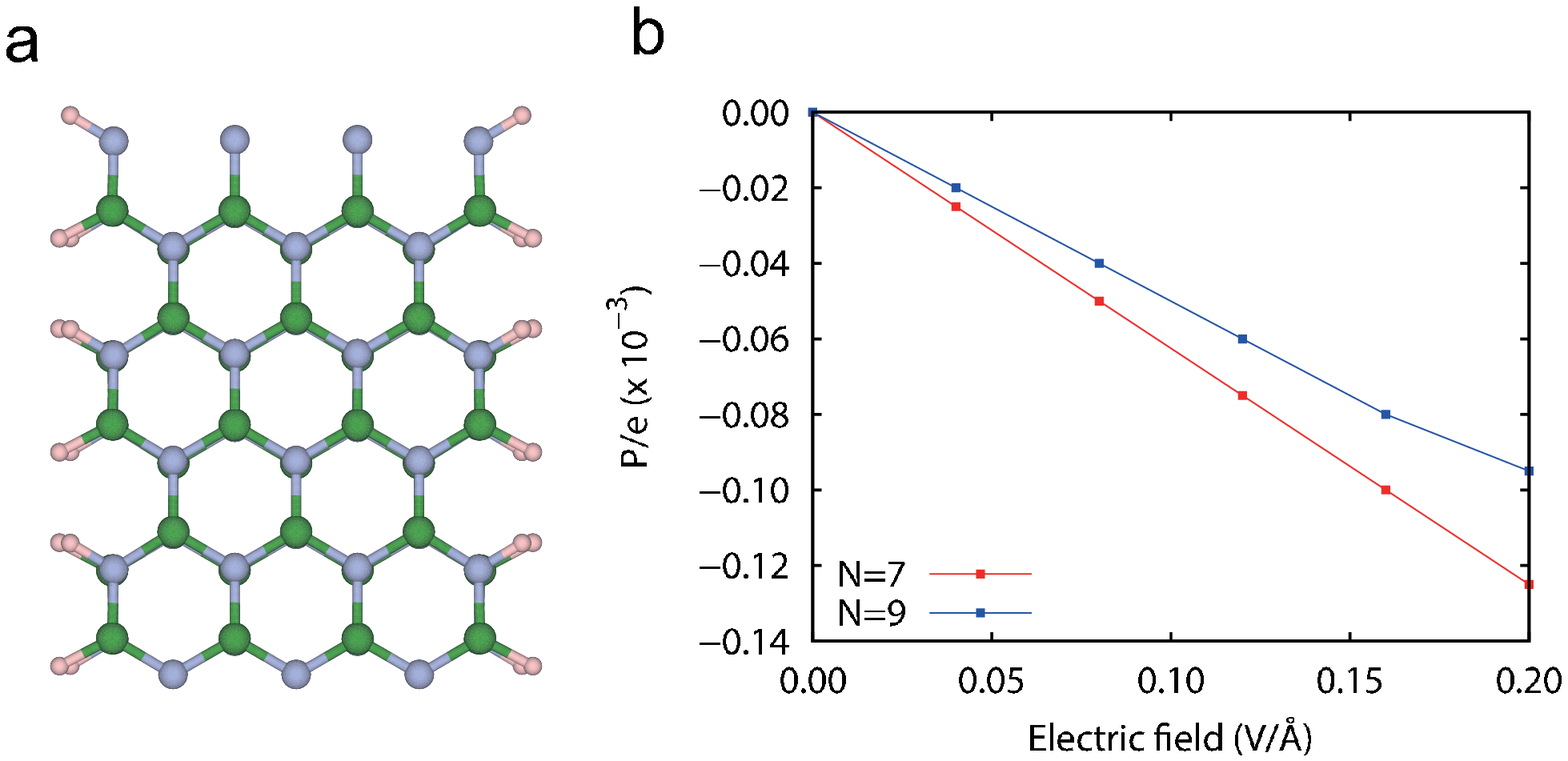}
\caption{\label{BNNR_polarization} \textbf{Polarization in the BN nanoribbons.}
\textbf{a} Top view of the hydrogen terminated bilayer BN nanoribbon with the armchair edges for $N=7$. 
B (N) atoms on the lower layer are just below the N (B) atoms on the upper layer.
\textbf{b} Electric field dependence of the polarization obtained by DFT calculations for bilayer BN nanoribbons with the armchair edges.
}
\end{figure}

\end{widetext}

\end{document}